# Development of Safety Monitoring System of Connected and Automated Vehicles Considering the Trade-Off between Communication Efficiency and Data Reliability

Sehyun Tak, and Seongjin Choi

*Abstract*—The safety of urban transportation systems is considered a public health issue worldwide, and many researchers have contributed to improving it. Connected automated vehicles (CAVs) and cooperative intelligent transportation systems (C-ITSs) are considered solutions to ensure urban transportation systems' safety using various sensors and communication devices. However, deploying the C-ITS framework in South Korea has been challenging because CAVs produce a massive amount of data every minute, but it cannot be transmitted via existing communication networks. Thus, raw data must be sampled for size reduction over a communication network and transmitted to the server for further processing. Moreover, the sampled data must be highly accurate to ensure the safety of different agents in C-ITS. Thus, we designed and developed a C-ITS architecture and data flow, including messages and protocols for the safety monitoring system of CAVs, and determined the optimal sampling interval for data transmission while considering the trade-off between communication efficiency and reliability of safety performance indicators. Three safety performance indicators were introduced: severe deceleration, lateral position variance, and inverse time to collision. A field test was conducted to collect data from various sensors installed in the CAV, determining the optimal sampling interval. Kolmogorov-Smirnov test was conducted to ensure statistical consistency between sampled and raw datasets. The effects of the sampling interval on message delay, data accuracy, and communication efficiency were analyzed. Consequently, a sampling interval of 0.2 s is recommended for optimizing the overall efficiency of the system.

## I. Introduction

THE safety of urban transportation systems has been recognized as a public health issue worldwide; consequently, many researchers have contributed to improving the safety performance. According to previous studies [1, 2], approximately 90 % of traffic incidents and accidents are caused by human factors. Therefore, many previous studies have used sensors and communication technologies to reduce the influence of human factors on traffic safety [3]. Although the researchers have focused on technologies using in-vehicle sensors, there exist several limitations in perception capability, detection range, and price in particular cases. Moreover, recent studies have concluded that connected and automated vehicles (CAVs) and cooperative intelligent transportation systems (C-ITS) can be a solution to ensuring transportation safety by reducing the perception and reaction error related to human drivers [4]. CAVs and C-ITS are based on information and communication technologies (ICT), which collect massive amounts of data from various sensors, transmit data and messages through telecommunications, process the data and messages to monitor real-time safety performance, and apply control strategies to vehicles and traffic management systems [5].

In the era of CAVs and C-ITS, there have been several changes in the role played by the transportation management center (TMC). First, traffic management is undergoing a shift from *infrastructure*- to *in-vehicle device*-based management that rely on mobile devices and/or in-vehicle communication units. In particular, previous research suggests that providing personalized traffic management can improve both traffic efficiency and safety compared to the conventional traffic management system, which provides traffic information based on the average behavior of road users [6,7,8,9,10]. Second, communication protocols in traffic management is witnessing a change from wired and/or mobile communications to vehicle-to-everything (V2X) and machine-to-machine (M2M) communications to ensure accurate and low-latency safety services. In the case of variable speed limit (VSL) and ramp metering, which are representative examples of conventional traffic management services, traffic control was performed using predefined schedules or transmitting a control method calculated in the server to the variable message signs (VMS) or traffic signals on the road through wired or mobile communication. However, in-vehicle device-based traffic management, using V2X or M2M communication is more efficient than the conventional communication method because

This paragraph of the first footnote will contain the date on which you submitted your paper for review. This research was supported by the Ministry of Land, Infrastructure, and Transport (MOLIT, Korea) under the Connected and Automated Public Transport Innovation National R&D Project (Grant no. 21TLRP-B146733-04).

Sehyun Tak is with Korea Transport Institute, 370 Sicheong-daero, Sejong-si, Republic of Korea

Seongjin Choi is with the Department of Civil and Environmental Engineering, Korea Advanced Institute of Science and Technology, Daejeon, Republic of Korea (e-mail: benchoi93@kaist.ac.kr)

it can provide the warning information to vehicle with low latency [11]. Finally, the nature of computing is changing from on-premise to cloud and/or edge computing to facilitate efficient and flexible data handling of multi-source unstructured big data. As mentioned previously, to provide in-vehicle device-based traffic management in real time, various computing tasks such as data acquisition, data preprocessing, and control decisions must be performed efficiently. Therefore, it would be ideal to subdivide computing roles (data acquisition, data preprocessing, communication, and data analytics) to cloud and edge rather than collecting and processing all data in one server as in on-premise computing.

Thus, to properly monitor the safety of urban transportation networks and apply proper traffic control, developing a novel framework design including data collection in vehicle devices, transmission from vehicles to cloud servers, and data processing in cloud platforms is vital. However, a problem encountered in developing and deploying such a framework is that the automated driving system produces a massive amount of data every minute. The size of the produced data varies from 1 to 3 *GB* per minute, and these data consist of data points collected by various sensors such as lidar, radar, vision, chassis, and GPS. According to a previous study on the communication speed in South Korea, the average data rate of LTE communication is 23.6 Mbps, which is equivalent to approximately 1.4 *GB* per minute [12]. In addition, the maximum data rate of WAVE communication is 27 *Mbps*, which is equivalent to approximately 1.6 *GB* per minute [13,14,15]. Thus, both existing communication systems cannot guarantee sufficient data rates to transmit data from connected automated vehicles in real time for safety applications.

In the near future, ensuring higher-speed communication for C-ITS may be possible due to the emergence of 5G and 6G communication systems. However, considering the current deployment stage, the size of data must be reduced through sampling to ensure that the data can be transmitted through the current communication system. On the other hand, it is necessary to acquire highly accurate data to ensure the safety of different agents in C-ITS. Such a trade-off relationship in traffic data has also been studied in previous studies focusing on the legacy intelligent transportation system [16,17,18]. Thus, both communication efficiency and data reliability in data collection must be considered for new real-time safety applications to properly monitor the safety performance of connected vehicles (CVs) and automated vehicles (AVs).

From research papers to standards, extensive studies have been undertaken related to frameworks and architectures for future transportation management centers. In particular, many studies have focused on C-ITS, which enables direct communication between vehicles, roadside infrastructure (also known as roadside units, RSUs), and TMCs. A high-level system architecture was proposed in [19], comprising end-to-end structure including a traffic control center, roadside infrastructure, and vehicle. In [20], an architecture for a cloud-based vehicle information system was proposed for C-ITS application services in connected vehicle environments. Use of the cloud system in C-ITS was proposed to process and store a massive amount of data generated from various sensors in C-ITS. Different types of data can be collected and analyzed under it, such as traffic flow and link speed data [21], safety-related data [22], and emission data [23]. With the emergence of CVs and CAVs, every vehicle traveling in a traffic network can function as a moving data generator and possibly merge with C-ITS. Furthermore, in [24], the exchange traffic data, traffic control data, and traffic management strategies between public authorities and the private sector were analyzed with possible levels of cooperation under Traffic Management 2.0 (TM2.0). The exchange of data collected by sensors under the control of C-ITS and data collected by sensors installed in CV and CAV are the key to aid in improvement of the overall performance of the traffic network. Although previous researches covered extensive ranges of studies with different aspects of C-ITS, there are many research gaps for real-world deployment. The previous studies focused on developing the conceptual design of C-ITS, and as a result, it is difficult to use them directly. Also, previous studies did not consider the massive amount of data that CAVs will produce when they emerge in urban transportation system. Therefore, it is necessary to fill the research gap to properly design and deploy C-ITS for CAVs on currently running communication networks by considering the trade-off between communication efficiency and data reliability.

This study primarily focused on the safety performance of an urban transportation network with CAV and C-ITS. Safety applications and safety-related decisions can be handled by edge devices installed in CAVs [25]; however, monitoring the safety performance of the overall network and confronting possible fallbacks of the automated driving system of CAVs are the roles that are undertaken by traffic management centers. Certain studies [22,26,27] have proposed a TMC-based collision warning system that includes data flow; however, they did not consider the communication efficiency, which may not be realistic at the deployment stage. Moreover, although the trade-off between communication efficiency through data sampling and data reliability has been studied in various fields, it has not been attempted in the context of CAV and C-ITS. In [18], a simplified resource allocation problem was defined to assess the trade-off between communication efficiency and data reliability under different sampling rate over a discrete-time communication channel. Furthermore, [16] analyzed the effects of different variables, such as the penetration rate and sampling frequency of mobile sensors, on traffic state estimation in a highway scenario.

Thus, the objectives of this study are 1) to develop and design a C-ITS architecture and data flow for the safety monitoring system of CAVs, and 2) to determine the optimal sampling interval for data transmission considering the trade-off between communication efficiency and data reliability of safety performance indicators. This study especially focuses on the design and deployment of C-ITS for CAVs considering the trade-off of data reliability and communication efficiency.

The remainder of this paper is presented as follows. In Section 2, we present a novel framework for traffic management centers for the safety monitoring of connected automated vehicles. In Section 2.A, we present the system architecture of the proposed traffic management center with a detailed explanation of the data flow for the safety monitoring system of CAVs. In Section 2.B, we present the safety performance indices used in this study. In Section 2.B, we present an efficient data transmission technology from an on-board unit to a cloud server. In Section 3, the data used in this study and the experimental design to determine the optimal sampling transmission interval are introduced. In Section 4, the results of applying various sampling intervals for the three safety performance indicators are discussed. Finally, in Section 5, the contributions of this study and future study plans are presented.

## II. METHODOLOGY

This study designed data flow for TMC based on data collected from CAVs for safety monitoring. The data flow ranges from the data generation in sensors in the CAVs to calculation of safety performance in TMC. In particular, data sampling method for transmitting the data collected from the in-vehicle sensors to TMC via Road Side Unit (RSU) is proposed. The details are described in the subsequent subsections.

### A. System Architecture

This study devises safety monitoring for CAVs, as shown in Figure 1. The proposed monitoring system is largely composed of four systems: CAD system, on-board unit (OBUs) for communication, roadside units (RSUs) for communication, and TMCs for CAVs. The CAD system controls the longitudinal and lateral movements of vehicles based on the information of the sensors installed in the vehicle. In addition, the data from the in-vehicle sensors such as chassis, vision sensor, radar sensor, lidar sensor, and GPS are first collected through the middleware such as the Robot Operation System (ROS). Each sensor has a different data collection cycle of less than 0.1 s, and the collected data are sent to the OBU for communication to be sent to the TMC for CAV.

The OBU sends the data collected from the CAD to the RSU. Because the data collected from the CAD comprises heterogeneous time intervals of various sensors, the unification of the heterogeneous time intervals from various sensors is necessary before sending the data to the RSU. Furthermore, data reduction is required due to the large size of raw data from the in-vehicle sensors that have to be sent over communication with limited transmission capability. Therefore, the OBU performs the sampling to unify the data received from the CAD at the same time intervals. For example, if the time interval of raw data is 0.01 s, the data are converted to 0.1 s units via capturing the raw data once every 10 points during the sampling work. Moreover, the data volume is also reduced by approximately 10 % in this process, although the accuracy (reliability) of the data was also reduced. Subsequently, the sampled data are packaged into the communication message format (V2X message) to be sent to the RSU [28], which is eventually sent to the TMC. Simultaneously, the communication state information of the RSU is generated and sent to the TMC.

The TMC performs functions to facilitate safety monitoring of CAVs based on the data received from the RSU, which are largely composed of the following three steps. First, the V2X message from the RSU are converted to a format that enables easier analysis in a server such as Jason, and then stored in the message broker, which can easily store and share data in real time. Second, a level of safety is calculated for each element of CAD to monitor the safety of the CAV based on the data stored in the message broker. This study used safety indicators for CAVs, such as severe deceleration, lateral position variation, and inverse time-to-collision. The safety indicators used are explained in the next section. Finally, the level of safety calculated through the safety indicators are displayed on the bulletin board to provide related information to the TMC manager, and the safety index for each vehicle/road section was calculated and used for vehicle maintenance, performance improvement, and warning of dangerous zones.

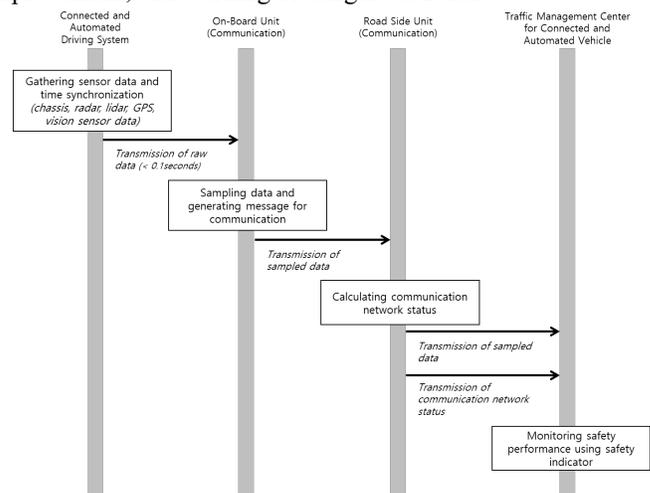

Figure 1. Data flow for Safety Monitoring of Connected and Automated Vehicle

### B. Indicators for monitoring of automated driving vehicle

This study uses three safety indicators for monitoring different aspects of safety: severe deceleration (SD), lateral position variation (LPV), and inverse time to collision (ITTC). The SD is designed to capture unsafe and uncomfortable longitudinal driving movements and is among the extensively used safety surrogate measures for indicating the safety performance of both drivers and CAVs [29, 30, 31]. The occurrence of SD is determined by the existence or absence of the following condition

$$Acc_{long}^{sub}(t) \leq D_{severe} \quad (1)$$

where $Acc_{long}^{sub}(t)$ is the longitudinal acceleration of subject vehicle at time t, and $D_{severe}$ is the threshold value for SD, which is set to $-2.94 \, m/s^2$ in this study.

This index comprehensively indicates the decreased comfort of users in the vehicle, the occurrence of an accident, and delayed cognition behavior of the preceding vehicle when there is a rapid deceleration of the vehicle. This is more strictly applied to mass transit, such as buses, due to the fact that safety of standing passengers in buses can be greatly decreased by the SD.

LPV represents vehicle safety related to lateral movement and is designed to capture the unsafe and unstable steering control of the lane-keeping system of connected and automated vehicles. The equation for LPV is as follows:

$$I_{LPV} = min\left(\left|y_{Left}^{dist} - \frac{l_{width}^{sub}}{2}\right|, \left|y_{Right}^{dist} - \frac{l_{width}^{sub}}{2}\right|\right) \quad (2)$$

where $y_{Left}^{dist}$ is the distance from the vehicle center to the left lane in the vehicle moving direction, $y_{Right}^{dist}$ is the distance from the vehicle center to the right lane in the vehicle moving direction, and $l_{width}^{sub}$ is the width of the subject vehicle.

The went over event, wherein a CAV approaches a lane too closely or crosses over a lane during driving, is a highly dangerous situation, and among the most widely used evaluation factors in field tests [32,33,34]. Use of LPV has been suggested to numerically analyze such dangerous events, and is considered as a representative key performance indicator for evaluating the driving safety of CAVs because it can detect a dangerous situation before the occurrence of a touched line event or went over event [35, 36].

The LPV determines whether the subject vehicle drives excessively to the left or right from the centerline. This implies that the minimum distance from the left or right side to the left or right lane, and the closer to zero this distance is, the more dangerous it is. For example, let us assume that a vehicle with a 2 m width is driving 0.5 m toward the right on a road with a lane width of 3.4 m. In this situation, $l_{width}^{sub}$ is 2 m, $y_{Left}^{dist}$ is $\frac{3.2m}{2} + 0.5$, and $y_{Right}^{dist}$ is $\frac{3.2m}{2} - 0.5$. Consequently, we obtain $I_{LPV} = min\left(\left|2.1 - \frac{2}{2}\right|, \left|1.1 - \frac{2}{2}\right|\right) = min(1.11, 0.11) = 0.11$. In this example, the vehicle is in a dangerous situation with a residual space of 0.11 m in the right direction. Consequently, the CAV must execute longitudinal vehicle control to return to the centerline or closely observe neighboring vehicles that approach from a close distance.

ITTC represents the collision risk of the subject vehicle considering the distance and relative velocity between the subject and the preceding vehicles [37, 38, 39, 40]. This indicator is often used to estimate the danger of the driver and CAV. The driver is provided a warning message to prevent an accident, and the CAV performs control such as acceleration and braking based on this value to maintain a safe distance. This study used the ITTC to evaluate the ability of CAVs to recognize and appropriately respond to objects in front. The equation for the ITTC is as follows:

$$I_{TTC}(t) = \frac{v_{sub}(t) - v_{pre}(t)}{d(t)} \quad (3)$$

where $v_{sub}(t)$ is the speed of the subject vehicle at time t, $v_{pre}(t)$ is the speed of the preceding vehicle at time t, and $d(t)$ is the distance between the rear of the preceding vehicle and the front of the subject vehicle.

An ITTC value closer to 0 indicates a safer situation, and as it attains larger positive values, a more dangerous situation is indicated Previous studies that used ITTC as a measure usually set the threshold to indicate a dangerous situation as 0.49 [37, 40]. However, in this study, we deal with the safety performance indicators of CAVs, and thus, a threshold same as human drivers should not be used. Considering previous studies, this study determined that an ITTC value larger than 1.76 is a dangerous situation [41, 42].

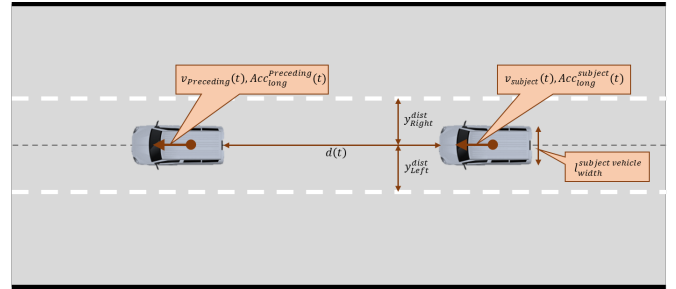

Figure 2. Explanation of variables used to calculate the safety indicators

### C. Sampling for transmission of automated driving data

As previously discussed, the amount of data produced by CAV every minute is approximately 1-3 GB. Under the current deployment stage, transmitting such large data is not possible via the existing communication networks in Korea, including WAVE (1.6 *GB/min*) and LTE (1.4 *GB/min*). Thus, as both communication systems cannot guarantee sufficient data rates to transmit data from connected automated vehicles in real time for safety applications, a sample data must be collected data before transmission through the communication network.

The collection frequency of the data varies depending on the sensors and target variables, ranging from 20 to 50 *Hz*. This implies that approximately 20 to 50 data points can be generated by the sensor every second. Thus, if one data point is sampled every second, theoretically, the size of the data can be reduced to one fiftieth, which can be accommodated by the current communication network for CAVs.

Adjusting the data sampling interval as explained above, can results in several benefits in terms of data transmission and storage, but the data accuracy decreases (error) and thus, a delay in the detection of severe events can occur. Moreover, it is possible that severe events may not have been detected in certain cases. Figure 3 presents examples of these problems. Figure 3 (a) shows the delay and error that occurred when a sampling interval of 10 *s* was applied to the data collected in a 20 *Hz* unit. This figure graphically shows the acceleration of the CAV. At this moment, the automated driving monitoring system must detect the SD situation. However, due to the occurrence of a delay because the system detected the peak point of SD slightly late, an error was produced that underestimated the severity of this event. Figure 3 (b) shows the result when a sampling interval of 5 *s* was applied to the data collected in a 20 Hz unit. In this situation, the automated driving monitoring system failed to detect SD events.

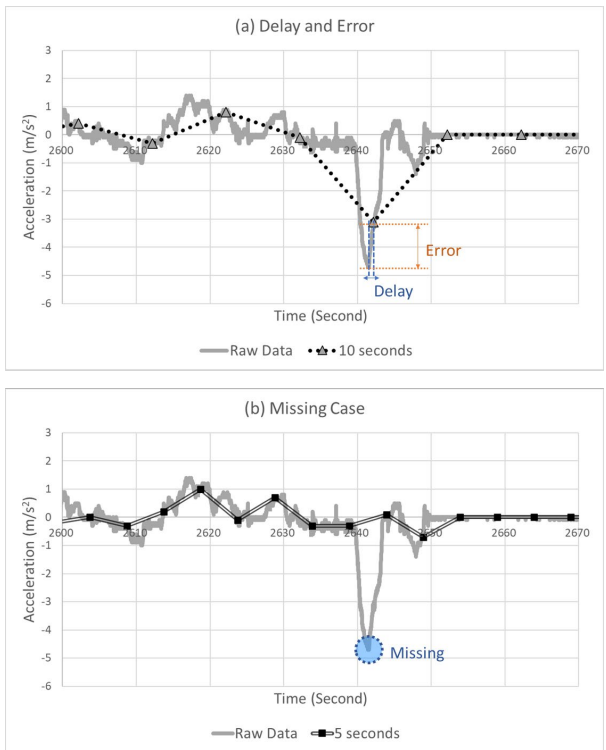

Figure 3. Example of (a) delay and error case and (b) missing case

The sampling interval may be increased, which reduces the data volume, and makes data communication and storage; however, as shown in Figure 3, an error and delay in severe event detection can occur, and severe events may not be detected in certain cases. In contrast, if the sampling interval is decreased, the data volume is increased, which imposes a large burden on data communication and storage. Thus, to satisfy these two requirements, a process for determining the optimal sampling interval is required.

Even if the sampled data are sent and stored by adjusting the sampling interval, if the data distribution is significantly changed compared to the raw data, the process becomes meaningless. Therefore, before determining the optimal sampling interval, this study first analyzed whether the distribution of the sampled data was similar to the distribution of raw data. Figure 4 shows examples of data distribution when data are acquired by applying a sampling interval as a probability density function. The black curve in each figure depicts the distribution of the values in the raw dataset, while the remaining curves depict the distribution of the values in the sampled dataset with different sampling intervals. The differences between the colored curves and the black curve in Figure 4 indicate that the distributions of data change by sampling data points from the raw data. Consequently, to ensure that this change of distribution does not cause statistical distributional change, we used the Kolmogorov-Smirnov test (KS test), an extensively used statistical test.

There are numerous options when choosing the correct statistical test, such as the t-test, Mann-Whitney test, and KS test [43]. In this study, we used the KS test, which is non-parametric, and does not make assumptions about or is not affected by the original distribution of data and is not affected by it [44]. The KS test quantifies the distance between the empirical cumulative distribution function of two given data. For two particular data points, the equation for the KS statistic is as follows:

$$D_{n,m} = \sup_{x}|F_{1,n}(x) - F_{2,m}(x)|$$

where $F_{1,n}$ and $F_{2,m}$ are the empirical distribution functions of the first and second samples, respectively, and $\sup$ is the supremum function.

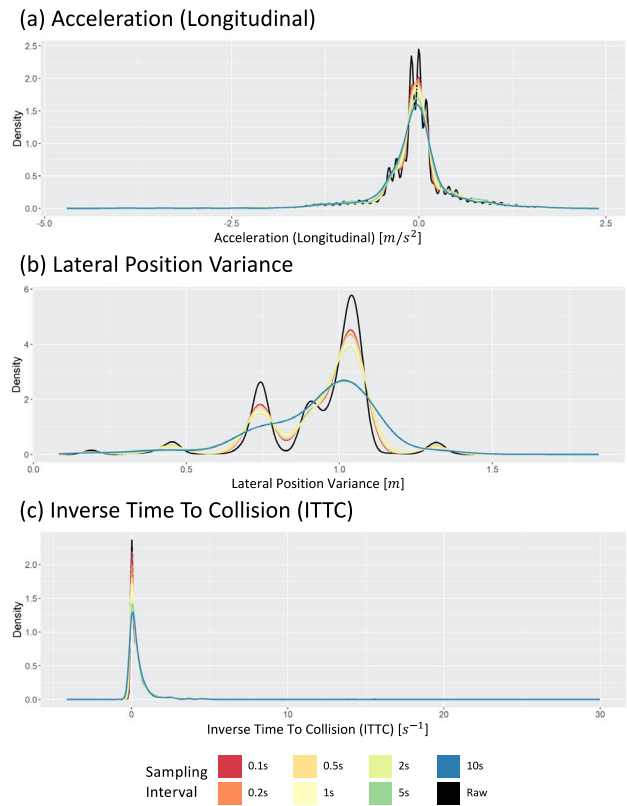

Figure 4. Changes in distribution of (a) Longitudinal Acceleration, (b) Lateral Position Variation, and (c) Inverse Time to Collision with different sampling intervals

Figure 5 shows the passing rate when the KS tests were performed 100 times. This result verifies whether the data obtained when sampling each data exhibit a distribution that is statistically similar to that of the raw data. It is evident that in the case of LPV, even a long sampling interval generates a statistically similar data distribution, but acceleration (longitudinal) cannot because the KS test passing rate drops after 10 s. Thus, various sampling intervals can be applied for LPV, but longitudinal acceleration can generate statistically different distributions when a sampling interval of 10 s is used. Furthermore, using a long sampling interval should be avoided because the ITTC data indicates a low passing rate, even when a short sampling interval is used.

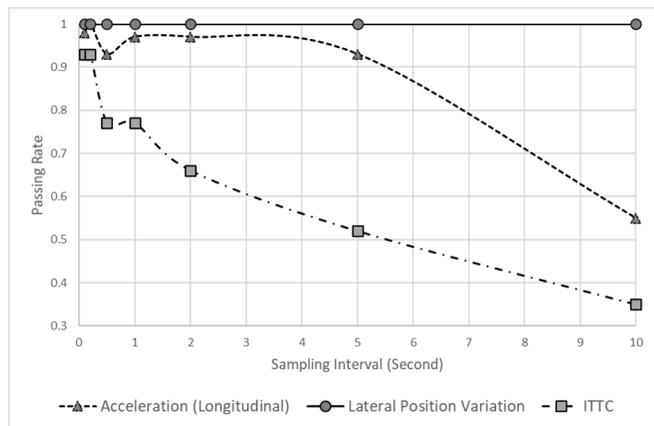

Figure 5. Passing Rate of Hypothesis with different Sampling Interval

## III. Data and Evaluation

To analyze the effect of sampling intervals on the reliability of indicators and communication efficiency, this study collected data from CAVs and adjusted the sampling intervals of the collected data. In the data collection stage, a vehicle with various sensors was driven on a real road to obtain data for various dangerous situations. Subsequently, the effect analysis stage, the effects of sampling intervals on the reliability of safety indicators and communication efficiency were analyzed. Furthermore, regarding the reliability of the safety indicator, the reduction in safety performance according to an increase in the sampling interval relative to the raw data was derived. The details are described in the following sections.

### A. Evaluation Metrics

As the sampling interval is increased, the performance is improved in terms of communication efficiency because the amount of collected/transmitted data is reduced. However, it causes a reduction in safety performance because safety-related events cannot be determined, as discussed in Section 2.C. Consequently, the safety performance in terms of the accurate detection of safety-related events and the communication efficiency in terms of the size of the data to be transmitted over communication exhibit a trade-off relationship.

The evaluation process of this study consists of three steps: analysis of sampled points, analysis of the delay and error of the sampled data, and evaluation of the overall objective. In the first step, we analyzed the characteristics of each safety performance indicator by analyzing the differences between the raw and sampled data points. In the second step, we analyzed the distribution of the error and delay of safety-related events. Finally, in the last step, we evaluated the overall performance of each sampling time interval based on the objective function considering both the communication efficiency and reliability of the safety indicator. We evaluate the communication efficiency in terms of the compression ratio *(how much the data size is reduced)* of the data, and evaluate the reliability in terms of detection success rate *(how many critical events are successfully detected)*. The objective function used in this study is as follows:

$$(w_{com} \cdot x^{ST=k}_{comp\ rate} + w_{rel} \cdot x^{ST=k}_{suc\ rate})$$

where $w_{com}$, $w_{rel}$ are weights for communication and reliability of the safety indicator, respectively, $x^{ST=k}_{zip\ rate}$ is the compression ratio of sampled data with sampling interval (k *s*) compared to raw data, and $x^{ST=k}_{suc\ rate}$ is the detection success rate of sampled data with sampling interval (k *s*) compared to raw data.

### B. Experimental Environment

Figure 6 shows the sensor configuration and dimensions of the CAVs used for data collection. The CAV has a length of 6.195 m, a width of 2.038 m, and a height of 2.665 m. Seven lidar sensors, two radar sensors, one vision sensor, and one GPS were installed in the CAV. The vision sensor collects image data from the vehicle and generates information related to objects and the environment, including the type of objects in front, the lane type, and the distance to the lane. Whereas, the radar sensors are used to detect objects in front and rear and generate information such as the distance and relative velocity to a vehicle that is relatively far away. Furthermore, lidar sensors generate information regarding the environment, such as safety facilities in the surroundings and information about objects such as people and vehicles. Finally, GPS generates information for estimating the location of vehicles and time synchronization. All information was collected, time-synchronized, and stored through the ROS.

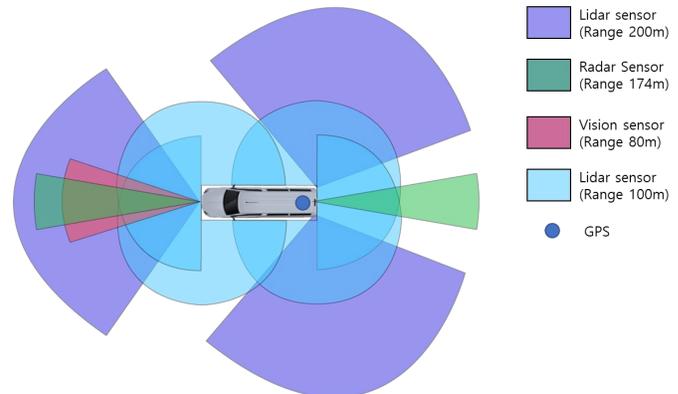

Figure 6 Sensor configuration of Connected and Automated Vehicle for experiment

Table 1 shows a description of the data collected by the connected and automated vehicles in this experiment. Three safety indicators were used in this study, as discussed in Section 2.B. SD is determined using the condition in Equation (1), which requires longitudinal acceleration data from the CAV. These data were collected using the Chassis sensor. The LPV is determined using the condition in Equation (2), which requires the position of the left lane and the position of the right lane. These data fields are collected by a vision camera sensor installed in front of the CAV. In addition, 'leftLaneQuality' and 'rightLaneQuality' were used to filter out noisy data. Finally, ITTC is determined using the condition in Equation (3), which requires the relative distance and relative speed between the ego vehicle and the obstacle in front of the CAV. These data were collected by the radar sensor installed in front of the CAV, and 'targetRangeAccel' (the relative acceleration) and 'targetStatus' were used to filter out noisy data.

TABLE 1. DESCRIPTION OF DATA COLLECTED BY CONNECTED AND AUTOMATED VEHICLE

| Data field | | Description |
|---|---|---|
| timestamp | | Data collection time (time synchronized through ROS) |
| Chassis | longAccel | Longitudinal acceleration of the ego vehicle. |
| Vision | leftLanePosition | position of left lane from the ego vehicle |
| | leftLaneQuality | quality of left lane detection |
| | rightLanePosition | Position of the right lane from the ego vehicle |
| | rightLaneQuality | Quality of right lane detection |
| Radar | targetRange | Distance between target and the ego vehicle |
| | targetRangeRate | Relative speed between target and the ego vehicle |
| | targetRangeAccel | Relative acceleration between target and the ego vehicle |
| | targetStatus | Status of the target |

The sensor data, including chassis, vision, and radar, were collected based on the test drives of CAVs conducted in Sejong city. We conducted eight test drives from August to November 2020, and the total driving time was 21 hours. One test drive was divided into two sessions: normal driving and abnormal driving. In the normal driving session, the CAV traveled one predefined route without causing any safety-related events. In contrast, in the abnormal driving session, the CAV traveled the pre-defined route six times, while generating safety-related events such as SD, left and right tilt, and rapid acceleration. The pre-defined route is shown in Figure 7, and a sample set of the collected data can be found at:
https://github.com/benchoi93/CAVTestDriveData.

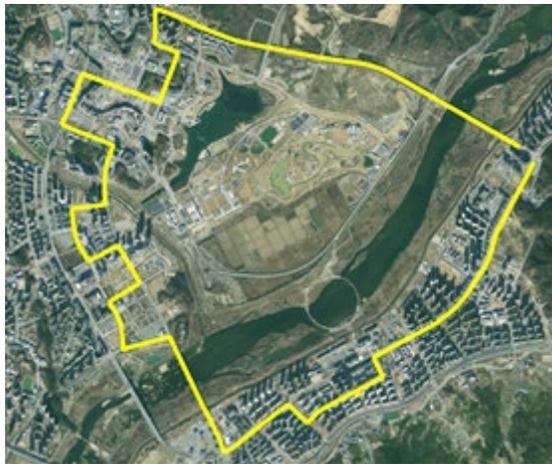

Figure 7 Pre-defined route for test drive in Sejong city (Skyview source: Kakao map, National Geographic Information Institute)

## IV. RESULTS AND ANALYSIS

### A. Severe Deceleration (Longitudinal)

Figure 8 shows an example of sampled points with different sampling intervals from 0.1 to 10 s. In the figure, the solid gray line and double black line represent the raw and sampled data, respectively. Furthermore, in the sampled data indicated by double black lines, the black squares represent the points where the data were collected. In the example case in Figure 8, the vehicle initially drove at a constant speed with an acceleration near 0, rapidly decelerated at up to – 3.4 m/s$^2$, and returned to the acceleration of 0 m/s$^2$. In this case, the amount of deceleration was increased for approximately 2 s, and subsequently, the amount of deceleration decreased for 10 s. The peak points occurred at 282.05 and 282.33 s. Moreover, as shown in Figure 8, as the sampling interval increased, the peak point of deceleration in the extracted data increased, which was delayed when compared with the raw data. The peak points of raw data occurred for the size of -3.4 m/s$^2$ at 282.05 and 282.33 s. However, as the sampling intervals increased to 0.1, 0.2, 0.5, 1, 2, 5, and 10 s, the peak points tended to increase to -3.3, -3.3, -3.3, -3.2, -2.8, -2.7, -2, and -0.9 m/s$^2$, respectively. Furthermore, the times when the peak points that occurred were delayed at 282.47, 282.51, 282.49, 283.25, 283.19, 284.69, and 289.85 s. Overall, as the sampling interval increased, the data smoothened, and the detection was delayed.

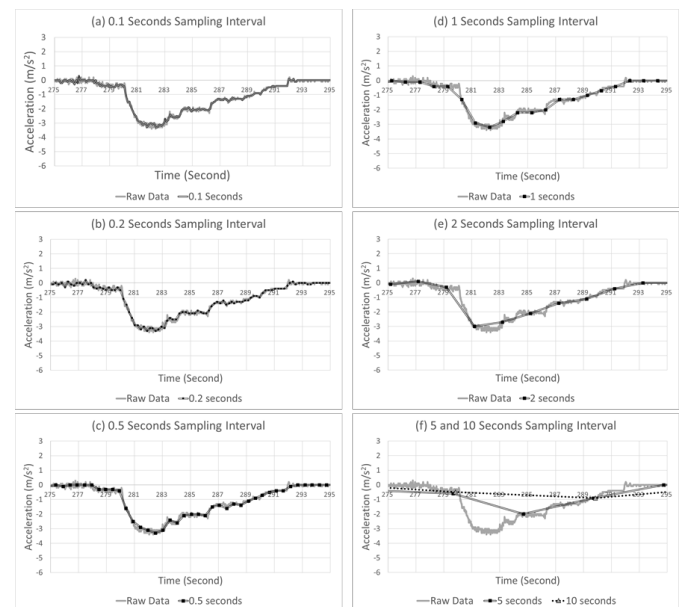

Figure 8 Example case of SD with different sampling interval

Figure 9 shows the changes in the distributions of delay and error at different sampling intervals. The figure shows the cases of detecting the occurrences of SD event (Figure 9(a), Figure 9(c)), and the cases of missing the occurrences of SD event (Figure 9(b), Figure 9(d)). Furthermore, Figure 9 (a) and (b) represent the time difference between the peak point of the raw data and the peak point of the sampled data. This time difference represents the delay of SD event detection. The increase of the sampling interval exhibits different effects on the distributions in detected case and missed case.

Figure 9(a) shows the delay distribution in the detected case. When the sampling interval increased from 0.1 to 10$s$, the mode of the delay increases from 0.088 to 0.172$s$, while the standard deviation increases from 0.041 to 1.30$s$. Thus, when the sampling interval increases, the increase in standard deviation is more clearly observed than the increase in the mode interval of delay. This is related to the duration of SD event, which is approximately 2.1$s$ on average. If the sampling interval is larger than the average of event duration, it is likely to miss the SD event.

Figures 9 (c) and (d) show the change in the error distribution of the sampled data at different sampling intervals. The effect of the sampling interval on the data distribution appears to be different in the detected and missed cases. In the detected case, when the sampling interval increases from 0.1 to 10$s$, the mode interval increases from 1.087 to 1.263 $m/s^2$ while, the standard deviation changes from 0.133 to 0.342 $m/s^2$. The change in standard deviation is larger than the change in the mode because the difference between the peak points of raw data and the threshold value of critical event (-2.94 $m/s^2$) is larger than the difference between the peak points of raw data and the peak points of sampled data. As a result, the detected cases become missed cases as the sampling interval increases. Therefore, the mode of the detected case is restricted by the peak point value of the raw data and threshold value. In contrast, in the missed case, as the sampling interval increases from 0.1 to 10 s, the mode increases from 0.329 to 3.040 $m/s^2$. In same situation, the standard deviation changes from 0.312 to 0.978 $m/s^2$, respectively. In the missed case, the changes in both the mode interval and standard deviation are large. In contrast to the detected case, in the missed case, the sampled data are smoothed toward 0 and the change of the sampled data are decreased when the sampling interval increases. Consequently, it becomes far from the mode and standard deviation of acceleration of the raw data.

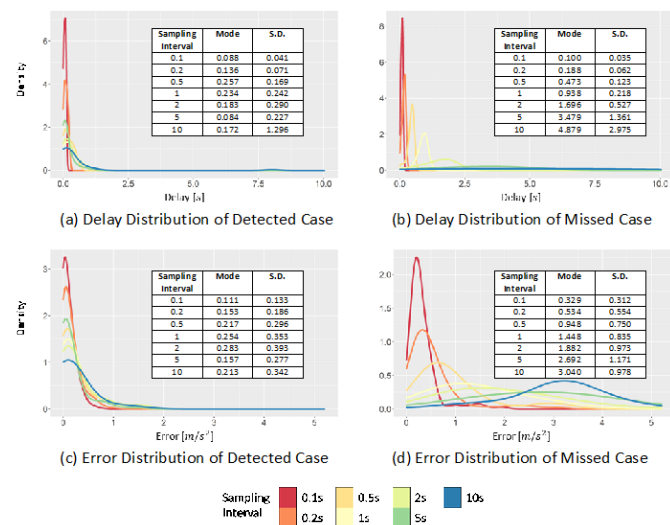

Figure 9. Changes in delay and error distribution of SD in detected case and missed case with different sampling interval

Figure 10 shows the changes in reliability and communication efficiency when different sampling intervals are applied. In this figure, reliability of indicators (in terms of detection success ratio) is marked by a solid black line and circles. When this value is closer to 1, it implies that all critical events are detected; whereas when closer to 0, fewer critical events than the raw data are detected. The communication efficiency (in terms of compression ratio) is marked by a dotted black line and black diamonds. This value is the ratio of the reduced amount of sampled data to the amount of raw data. The closer it is to 1, the greater the size of data reduction, and vice versa. Finally, the weighted average of the detection success and compression ratios is marked by a black dotted line and black triangles, and the weight was set to 0.5 each.

As shown in Figure 10, when the sampling interval increased from 0.1 to 5 s, the detection success ratio decreased. However, the degree of the decrease is different. When the sampling interval changes from 0 (raw data) to 1 s, the detection success ratio decreases by 0.454, but when it changed from 1 to 5 s, it decreases by 0.329. In contrast to the detection success ratio, the compression ratio continuously increases with the sampling interval. However, the increasing trend appeared differently depending on the sampling interval. When the sampling interval changes from 0 (raw data) to 0.2 s, the compression ratio increases by 0.905, but when it changes from 0.2 to 5 s, it increases by 0.091. Moreover, because of the different trends of the detection success and compression ratios, a peak point of the weighted sum of the two values was observed at 0.2 s. In other words, before 0.2 s, the rapid increase in the compression ratio has a large effect on the weighted sum, thus increasing the weighted sum. However, after 0.2 s, the effect of the detection success ratio was larger, thus decreasing the weighted sum. Considering these results, 0.2 s is recommended as the sampling interval for SD.

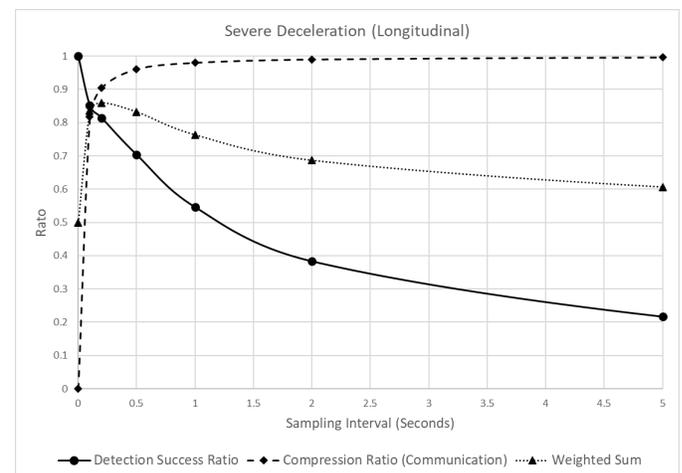

Figure 10. Analysis of detection success ratio, compression ratio, and weighted sum of SD with different sampling interval

B. *Lateral Position Variation (LPV)*

Figure 11 shows an example case of sampling points with different sampling intervals ranging from 0.1 to 10$s$. In this figure, the solid gray and double black lines indicate the raw and sampled data, respectively. In the sampled data, the data marked by black squares represent the points data collection points. The example case in Figure 11 shows a vehicle driving close to the right with an LPV of approximately 0.765*m* and further moving close to the right with an LPV of 0.190 m at approximately 95*s*. The margin to the right is approximately 0.190 m, which is a dangerous situation for lateral control leading to a "touched line event." In contrast to the case of SD in Figure 8, error and delay of LPV shows the different patterns. In terms of error, the peak point in all time sampling intervals is observed near 0.190. The differences between peak point of raw data and that of sampled data is very little. In contrast, the time at which a peak point occurred was delayed as the sampling interval increased; that is, the error experiences minimal change, but the delay increased to 0.100, 0.182, 0.420, 0.827, 1.619, 3.864, and 7.041*s* when the sampling interval increased to 0.1, 0.2, 0.5, 1, 2, 5, and 10*s*, respectively.

The error and delay show different patterns due to the duration of the event when a critical LPV appeared. In the case of SD in Figure 8, the duration is only 2.1 s, whereas the duration of a critical event of LPV is approximately 19.99 *s*. Therefore, in the case of LPV, even if the delay increases due to a long sampling interval, it does not affect the detection of a critical event itself.

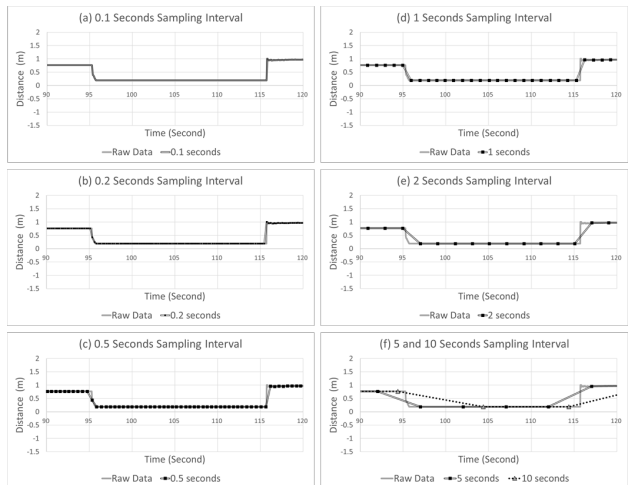
Figure 11 Example case of LPV with different sampling interval

Figure 12 shows the changes in the distributions of delay and error at different sampling interval. When the sampling interval increases from 0.1 to 10*s*, the mode interval of the delay of the detected case (Figure 12 (a)) increases from 0.16 to 8.57*s*, while the standard deviation also increases from 0.043 to 3.561*s*. This result is different from the result of delay distribution of SD of the detected case (Figure 9 (a)), where only the standard deviation changed while the mode underwent minimal change. In the case of LPV, the mode and standard deviation increase as the sampling interval increases because of the long duration of a critical event. In the detected case of LPV, the negative effect caused by increasing sampling interval is reflected in the distribution of delay. In addition, the effect of such a long duration was also observed in the missed cases. When the sampling interval increases from 0.1 to 10*s*, the mode interval of the delay distribution of missed cases (Figure 12 (b)) increases from 0.109 to 6.374*s*, while the standard deviation also increases from 0.030 to 2.661*s*. The increasing trends of both the mode interval and standard deviation can be clearly observed.

Due to the effect of the relatively long duration of the event, the effect of the sampling interval on the LPV error was not large. As shown in Figures 12(c) and (d), even when the sampling interval increased from 0.1 to 10 s, the mode interval of the error distribution of detected cases changed in small increments from 0.001 to 0.016*m*. In addition, the standard deviation also changes from 0.005 to 0.016*m*, which is the smallest of increments compared to the other indicators. Moreover, a similar trend was also observed in the error distribution of the missed cases. This is because the duration of the peak points of the critical event of LPV is often longer than the sampling interval, and even if the sampling interval increases, all dangerous situations have already been detected.

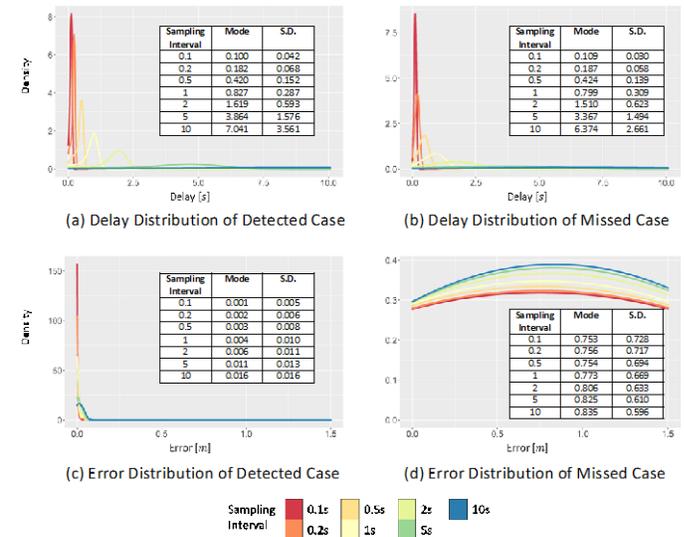
Figure 12. Changes in delay and error distribution of LPV in detected case and missed case with different sampling interval

Figure 13 shows the changes in the reliability of the LPV and the communication efficiency of the LPV when the sampling interval is increased. As evident, in contrast to the SD case (Figure 10), the effect of the sampling interval change on the detection success ratio of the LPV is relatively small. Thus, at a sampling interval of 0.1 s, the detection success ratio is 0.999, whereas at 10 s, it is 0.951. In contrast, the compression ratio increased significantly from 0.857 at 0.1 s to 0.974 at 10 s. Therefore, the sampling interval in the LPV has a much larger effect on the compression ratio than on the detection success ratio.

Owing to the contrasting effects of the sampling interval on the compression and detection success ratios, the time when the peak point of the weighted sum appears is also different from that of SD. The peak value of SD appeared when the sampling interval was 0.2, whereas that of LPV appeared at 5 s. Thus, in the case of LPV, even if a relatively long sampling interval is applied, dangerous situations can be detected more effectively than in SD.

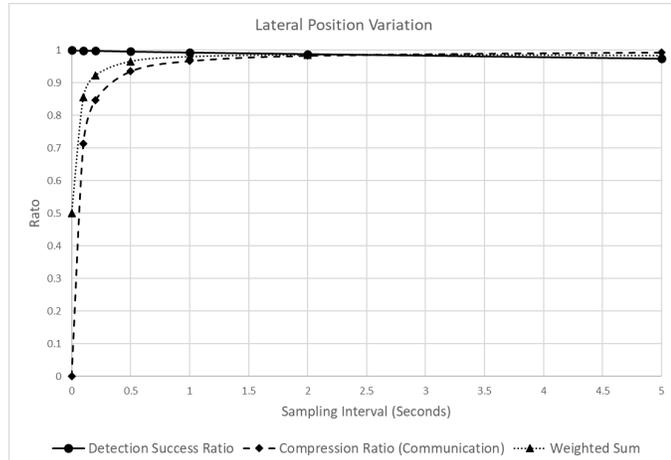

Figure 13. Analysis of detection success ratio, compression ratio, and weighted sum of LPV with different sampling interval

### C. Inverse Time to Collision (ITTC)

Figure 14 shows an example case of sampling points with different sampling intervals ranging from 0.1 to 10$s$. In the figure, the solid gray and double black lines represent the raw and sampled data, respectively. Furthermore, in the sampled data indicated by a double black line, the black squares represent the points the data collection points. The example case in Figure 14 shows a change in the ITTC. A critical event with an ITTC greater than $2.995s^{-1}$ occurred near 1718.88$s$ and another critical event with an ITTC of $0.987s^{-1}$ occurred near 1731.38$s$. Considering that in general, an ITTC value higher than $1.76s^{-1}$ is regarded as a critical event [41, 42], the example case includes an urgent situation where an accident can occur if a prompt response is not made.

When compared with the SD and LPV cases described earlier, the sampling interval of the ITTC had a significant effect on the detection accuracy. As shown in the 0.1-sec sampling interval of Figure 14 (a), even when it was slightly increased by 0.1 s from the average time interval of 0.0499$s$ for raw data, the dangerous situation of an ITTC of $2.995s^{-1}$, which occurred near 1718.88$s$, was not be detected. Furthermore, a slightly dangerous situation of an ITTC of $0.987s^{-1}$ at 1731.38 s, which maintained a relatively long duration, was barely detectable at sampling intervals longer than 1$s$. Furthermore, certain situations were not detected at the sampling intervals of 0.1 – 0.5$s$. Consequently, in contrast to the occurrence of a peak point value of $2.995\,s^{-1}$ at 1718.88$s$ in the raw data, when the sampling interval was 0.1, 0.2, 0.5, 1, 2, 5, and 10$s$, the peak points occurred at 0.987, 0.987, 0.987, 0.151, 0.127, 0.111, and $0.124s^{-1}$, respectively. Moreover, no dangerous situation is detected when the sampling interval is larger than 1 s.

The ITTC reacts more sensitively to the increase in the sampling interval than other indicators because the duration of the critical event of ITTC is the shortest among the indicators. In the example case, the duration of ITTC is around 0.1$s$. Therefore, it fails to capture the peak points even if the sampling intervals is slightly increased. Due to the short duration of ITTC, significant changes of ITTC value are not observed when the sampling interval is larger than 1s.

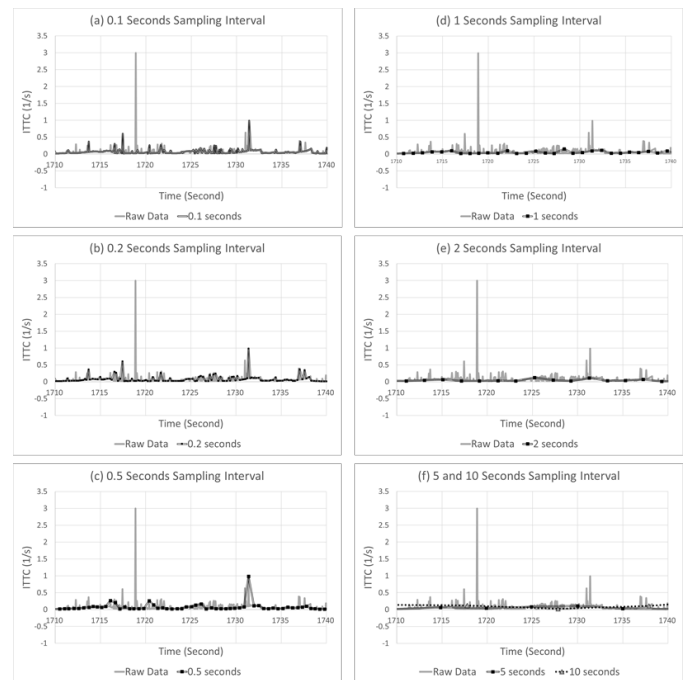

Figure 14 Example case of ITTC with different sampling interval

Figure 15 shows the changes in the delay and error distributions with different sampling intervals for the detected and missed cases. As shown in Figure 15 (a), in the delay distribution of detected cases, when the sampling interval increases from 0.1 to 10$s$, the mode interval increased significantly from 0.050 to 4.138$s$, and the standard deviation also increased significantly from 0.049 to 3.220$s$. Similar large increases in the mode interval and standard deviation were also observed in the delay distribution of the missed cases (Figure 15 (b)). Due to the large increase in the standard deviation according to the increase in the sampling interval, when the sampling interval is larger than 5 s, uniform distribution that is almost flat is obtained. The large increase occurs because when the sampling interval increases, the critical event in the raw data can be barely detected, rather than detecting it even at a later time. Large increases in the mode interval and standard deviation is also observed in Figures 15(c) and (d), indicating the error distribution changes with different sampling intervals.

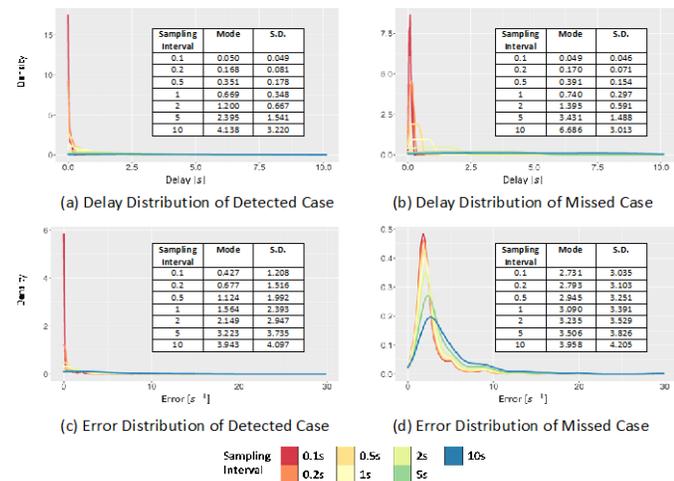

Figure 15. Changes in delay and error distribution of ITTC in detected case and missed case with different sampling interval

Figure 16 shows the changes in reliability and communication efficiency when the sampling interval increases. As shown in this figure, the detection success ratio decreases sharply when the sampling interval increases. In particular, when the sampling interval in raw data increases to 0.1, 0.2, and 0.5 s, the detection success ratio decreases significantly to 0.502, 0.606, and 0.746. Subsequently, the decrease becomes smaller, and when the sampling interval becomes 10 s, the change is close to zero. Furthermore, the compression ratio, which represents the communication efficiency, increases significantly by 0.905 till the sampling interval is 0.5 s, and then the change converges to 1, similar to the pattern of other indicators. Due to the opposite changes in the detection success and compression ratios, the weighted sum of the two values reaches a peak near 0.2 s. However, in contrast to the SD results (Figure 10), the changes in the weighted sum with different sampling intervals show minimal variations between 0.5 and 0.6. This is due to the opposite tendencies of the detection success and compression ratio, which change at similar values. This suggests that the effect of the sampling interval on the overall performance of the indicator is practically insignificant.

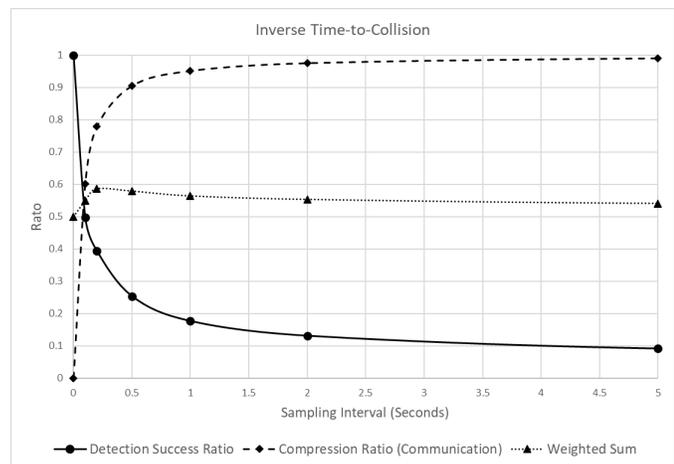

Figure 16. Analysis of detection success ratio, compression ratio, and weighted sum of ITTC with different sampling interval

## V. Conclusion

In this study, we propose a safety monitoring system for connected and automated vehicles and analyzed the effect of different sampling intervals on reliability and communication efficiency based on three indicators indicating various aspects of safe driving. The analysis results showed that the increase in the sampling interval has similar effects on the communication efficiency for all three indicators. Although the degree of change differs depending on the size of raw data required for indicator calculation, all three indicators show more than 90 % improvement in communication efficiency when the sampling interval is larger than 0.5 s. Regarding the accuracy of the indicators, different results are observed for the three indicators. In the case of SD, when the sampling interval is 0.5$s$, the success ratio was approximately 0.7, showing an intermediate effect on accuracy among the three indicators. In the case of LPV, the accuracy was 0.996 when the sampling interval was 0.5 s, and the success ratio changed very little because of the long duration of the critical event. In the case of ITTC, the success ratio changed the largest to 0.254 when the sampling interval was 0.5 s due to the short duration of the critical event. Because of the different change patterns of the success ratio of the three indicators, the optimal sampling interval for each indicator was derived as 0.2 sec for SD, 5 sec for LPV, and 0.2 sec for ITTC.

For efficient and reliable monitoring of the safety indicators, different sampling intervals should be applied to each indicator to obtain optimal results. However, as shown in the data flow for safety monitoring of CAV in Figure 1, the communication messages transmitted from the CAD to the RSU and the center are designed for integrated utilization of the safety monitoring of CAD, rather than for each indicator. Thus, designing and applying messages with different sampling intervals to the CAD monitoring system would probably decrease the integration efficiency due to the increased number of messages. Therefore, it is more advantageous to uniformly apply a sampling interval of 0.2 $s$ in terms of applying indicators with different characteristics. Furthermore, in the case of indicators with a long duration of the critical event, such as LPV, the increase in gain obtained by increasing the sampling interval is not large compared to the other 2 indicators. For example, when the sampling interval is changed from 5 sec to 0.2 sec, the decrease in the weighted sum of LPV is small at 0.061, whereas those of the SD and ITTC increase significantly by 0.253 and 0.045, respectively. To optimize the overall efficiency of the system in such heterogeneous features of safety indicators, a sampling interval of 0.2 s is recommended.

In this study, although the sampling interval for safety monitoring was researched based on three safety indicators with different characteristics, it was also observed that the sampling interval of 0.2 s suggested can be changed depending on the characteristics of the indicators. For example, if there are many indicators with a long duration, such as LPV in the overall safety monitoring system, the optimal sampling interval will be larger than the result of this study. Therefore, when communication messages for the safety monitoring of CAD are designed, not only the reduction in efficiency, but also the characteristics of the indicator should be considered. In the future, the characteristics of more diverse safety indicators will be analyzed by expanding the results of this study. Furthermore, although this study assumed the same sampling interval and data transmission interval, the follow-up paper will also analyze the optimal method for data transmission from connected and automated vehicles when different sampling intervals and transmission intervals are applied.

REFERENCES

[1] M. Behnood, and A. Pakgohar. "The Preliminary Plan for Budgeting the Costs of Social Training of Traffic." Organization of Transportation and Traffic Management, Tehran (2008).
[2] A. Pakgohar, et al., "The role of human factor in incidence and severity of road crashes based on the CART and LR regression: a data mining approach," *Procedia Computer Science*, vol. 3, pp. 764-769, 2011. DOI: https://doi.org/10.1016/j.procs.2010.12.126
[3] A. Sarker, et al., "A review of sensing and communication, human factors, and controller aspects for information-aware connected and automated vehicles," *IEEE Transactions on Intelligent Transportation Systems,* vol. 21, no. 1, pp. 7-29, 2019. DOI: https://doi.org/10.1109/TITS.2019.2892399
[4] L. Wang, et al., "How many crashes can connected vehicle and automated vehicle technologies prevent: A meta-analysis," *Accident Analysis & Prevention,* vol. 136, pp. 105299, 2020.
[5] M. Lu, et al., "C-ITS (cooperative intelligent transport systems) deployment in Europe: challenges and key findings," presented at the *25th ITS World Congress*, Copenhagen, Denmark, April 18-21, 2018.
[6] C. Vallon, et al., "A machine learning approach for personalized autonomous lane change initiation and control," presented at *IEEE Intelligent Vehicles Symposium (IV)*. Los Angeles, CA, USA, June 11-14, 2017.
[7] S. Choi, J. Kim and H. Yeo, "TrajGAIL: Generating urban vehicle trajectories using generative adversarial imitation learning", Transportation Research Part C: Emerging Technologies, vol. 128, p. 103091, 2021. DOI: https://doi.org/10.1016/j.trc.2021.103091
[8] S. Choi, H. Yeo and J. Kim, "Network-wide vehicle trajectory prediction in urban traffic networks using deep learning", Transportation Research Record: Journal of the Transportation Research Board, vol. 2672, no. 45, pp. 173-184, 2018. DOI: https://doi.org/10.1177/0361198118794735
[9] N. Bao, et al., "Personalized safety-focused control by minimizing subjective risk," presented at *2019 IEEE Intelligent Transportation Systems Conference (ITSC),* Auckland, New Zealand, Oct. 27-30, 2019.
[10] B. Zhu, et al., "Typical-driving-style-oriented personalized adaptive cruise control design based on human driving data," *Transportation Research Part C: Emerging Technologies*, vol. 100, pp. 274-288, 2019. DOI: https://doi.org/10.1016/j.trc.2019.01.025
[11] S. Nallamothu, et al., "Detailed Concept of Operations: Transportation Systems Management and Operations/Cooperative Driving Automation Use Cases and Scenarios," Federal Highway Administration, Office of Safety and Operations Research and Development McLean, VA, USA, FHWA-HRT-20-064, 2020.
[12] S. -T. Park, H. Im, and K. -S. Noh, "A study on factors affecting the adoption of LTE mobile communication service: The case of South Korea," *Wireless Personal Communications*, vol. 86, no. 1, pp. 217-237, 2016. DOI: https://doi.org/10.1007/s11277-015-2802-7
[13] D. Jiang and L. Delgrossi, "IEEE 802.11 p: Towards an international standard for wireless access in vehicular environments," presented at *VTC Spring 2008-IEEE Vehicular Technology Conference*, Marina Bay, Singapore, May 11-14, 2008.
[14] S. Eichler, "Performance evaluation of the IEEE 802.11 p WAVE communication standard," presented at *2007 IEEE 66th Vehicular Technology Conference*, Baltimore, MD, USA, Sept. 30 – Oct. 3, 2007.
[15] Z. H. Mir and F. Fethi, "LTE and IEEE 802.11 p for vehicular networking: a performance evaluation," *EURASIP Journal on Wireless Communications and Networking*, vol. 2014, no. 1, pp. 1-15, 2014. DOI: https://doi.org/10.1186/1687-1499-2014-89
[16] C. Bucknell and J. C. Herrera, "A trade-off analysis between penetration rate and sampling frequency of mobile sensors in traffic state estimation," *Transportation Research Part C: Emerging Technologies*, vol. 46, pp. 132-150, 2014. DOI: https://doi.org/10.1016/j.trc.2014.05.007
[17] J. Gao, et al., "Trade-off between sampling rate and resolution: A time-synchronized based multi-resolution structure for ultra-fast acquisition," presented at *2018 IEEE International Symposium on Circuits and Systems (ISCAS)*, Florence, Italy, May 27-30, 2018.
[18] C. Zhang and X. Liu, "Trade-off between the sampling rate and the data accuracy," presented at *American Control Conference*, Seattle, WA, USA, June 11-13, 2008
[19] A. Paier, "The end-to-end intelligent transport system (ITS) concept in the context of the European cooperative ITS corridor." 2015 IEEE MTT-S International Conference on Microwaves for Intelligent Mobility (ICMIM). IEEE, 2015.
[20] S. M. Song, et al., "A Prospective Cloud-Connected Vehicle Information System for C-ITS Application Services in Connected Vehicle Environment," presented at *International Conference on Cloud Computing. Springer*, Cham, 2015.
[21] K. Zhang, et al., "Hybrid short-term traffic forecasting architecture and mechanisms for reservation-based Cooperative ITS," *Journal of Systems Architecture*, vol. 117, pp. 102101, 2021. DOI: https://doi.org/10.1016/j.sysarc.2021.102101
[22] S. Tak, et al., "Sectional information-based collision warning system using roadside unit aggregated connected-vehicle information for a cooperative intelligent transport system," *Journal of advanced transportation*, vol. 2020, 2020. DOI: https://doi.org/10.1155/2020/1528028
[23] H. Vojvodić, M. Vujić, and P. Škorput, "Cooperative architecture of data acquisition for emission reduction in traffic," presented at *2017 25th Telecommunication Forum (TELFOR)*, Belgrade, Serbia, Nov. 21-22 2017.
[24] T. Vlemmings, et al., "Contractual Agreements in Interactive Traffic Management–looking for the optimal cooperation of stakeholders within the TM 2.0 concept," proceedings ITS European Congress 2017. 2017.
[25] M. A. Javed, S. Zeadally, and E. B. Hamida, "Data analytics for cooperative intelligent transport systems," *Vehicular Communications*, vol. 15 pp. 63-72, 2019.
[26] D. Lee, et al., "Real-time feed-forward neural network-based forward collision warning system under cloud communication environment," *IEEE Transactions on Intelligent Transportation Systems*, vol. 20, no. 12, pp. 4390-4404, 2018. DOI: https://doi.org/10.1109/TITS.2018.2884570
[27] S. Tak, S. Woo, and H. Yeo, "Study on the framework of hybrid collision warning system using loop detectors and vehicle information," *Transportation Research Part C: Emerging Technologies,* vol. 73, pp. 202-218, 2016. DOI: https://doi.org/10.1016/j.trc.2016.10.014
[28] R. Sumner, B. Eisenhart, and J. Baker, "SAE J2735 standard: applying the systems engineering process," Department of Transportation, Intelligent Transportation Systems Joint Program Office, Washington, DC, USA, FHWA-JPO-13-046, Jan. 2013.
[29] N. J. Goodall and C. -L. Lan, "Car-following characteristics of adaptive cruise control from empirical data," *Journal of Transportation Engineering, Part A: Systems*, vol. 146 no. 9, pp. 04020097, 2020. DOI: https://doi.org/10.1061/JTEPBS.0000427
[30] G. I. Jin, et al., "Experimental validation of connected automated vehicle design among human-driven vehicles," *Transportation Research Part C: Emerging Technologies,* vol. 91, pp. 335-352, 2018. DOI: https://doi.org/10.1016/j.trc.2018.04.005
[31] G. Orosz, et al., "Seeing beyond the line of site-controlling connected automated vehicles," *Mechanical Engineering*, vol. 139, no. 12, pp. S8-S12, 2017. DOI: https://doi.org/10.1115/1.2017-Dec-8
[32] M. Kim, et al., "Vision-based uncertainty-aware lane keeping strategy using deep reinforcement learning," *Journal of Dynamic Systems, Measurement, and Control*, vol. 143 no. 8, pp. 084503, 2021. DOI: https://doi.org/10.1115/1.4050396
[33] J. B. Cicchino, "Effects of lane departure warning on police-reported crash rates," *Journal of Safety Research*, vol. 66, pp. 61-70, 2018. DOI: https://doi.org/10.1016/j.jsr.2018.05.006
[34] W. Wang, et al., "A learning-based approach for lane departure warning systems with a personalized driver model," *IEEE Transactions on Vehicular Technology*, vol. 67, no. 10, pp. 9145-9157, 2018. DOI: https://doi.org/10.1109/TVT.2018.2854406

[35] L. Wang, et al., "How many crashes can connected vehicle and automated vehicle technologies prevent: A meta-analysis," *Accident Analysis and Prevention*, vol. 136, pp. 105299, 2020. DOI: https://doi.org/10.1016/j.aap.2019.105299

[36] S. Sternlund, et al., "The effectiveness of lane departure warning systems—A reduction in real-world passenger car injury crashes," *Traffic Injury Prevention*, vol. 18, no. 2, pp. 225-229, 2017. DOI: https://doi.org/10.1080/15389588.2016.1230672

[37] H. Yu, et al., "Impact of autonomous-vehicle-only lanes in mixed traffic conditions," *Transportation Research Record*, vol. 2673, no. 9, pp. 430-439, 2019. DOI: https://doi.org/10.1177%2F0361198119847475

[38] R. J. Kiefer, D. J. LeBlanc, and C. A. Flannagan, "Developing an inverse time-to-collision crash alert timing approach based on drivers' last-second braking and steering judgments," *Accident Analysis and Prevention*, vol. 37 no. 2, pp. 295-303, 2005. DOI: https://doi.org/10.1016/j.aap.2004.09.003

[39] V. E. Balas and M. M. Balas, "Driver assisting by inverse time to collision," presented at *2006 World Automation Congress*, Budapest, Hungary, Jul. 24-26, 2006.

[40] S. Moon, I. Moon, and K. Yi, "Design, tuning, and evaluation of a full-range adaptive cruise control system with collision avoidance," *Control Engineering Practice*, vol. 17, no. 4, pp. 442-455, 2009. DOI: https://doi.org/10.1016/j.conengprac.2008.09.006

[41] J. Khoury, K. Amine, and R. A. Saad, "An initial investigation of the effects of a fully automated vehicle fleet on geometric design," *Journal of Advanced Transportation*, vol. 2019, 2019. DOI: https://doi.org/10.1155/2019/6126408

[42] V. V. Dixit, S. Chand, and D. J. Nair, "Autonomous vehicles: disengagements, accidents and reaction times," *PLoS One*, vol. 11, no. 12, pp. e0168054, 2016. DOI: https://doi.org/10.1371/journal.pone.0168054

[43] B. K. Nayak, and A. Hazra, "How to choose the right statistical test?," *Indian Journal of Ophthalmology*, vol. 59, vol. 2, pp. 85, 2011. DOI: https://doi.org/10.4103/0301-4738.77005

[44] D. S. Dimitrova, V. K. Kaishev, and S. Tan, "Computing the Kolmogorov-Smirnov distribution when the underlying CDF is purely discrete, mixed, or continuous," *Journal of Statistical Software*, vol. 95, no. 1, pp. 1-42, 2020. DOI: http://doi.org/10.18637/jss.v095.i10

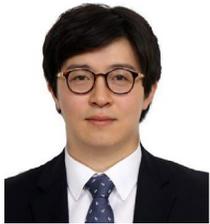

**Sehyun Tak** was born in Seoul, South Korea, in 1982. He received his M.S. and Ph.D. degrees in civil and environmental engineering from KAIST, Daejeon, South Korea, in 2011 and 2015, respectively. He is currently an Assistance Research Fellow at the Korea Transport Institute. His current research interests include connected and automated vehicles, shared mobility, C-ITS, and cloud-based big data analysis.

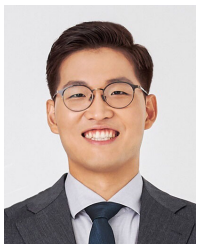

**Seongjin Choi** was born in Seoul, South Korea in 1993. He received the B.S., M.S., and Ph.D. degree in Civil and Environmental Engineering from Korea Advanced Institute of Science and Technology (KAIST), Daejeon, South Korea, in 2015, 2017, and 2021, respectively. He is a postdoctoral researcher at KAIST at AIxMobility Lab. His research interests include urban mobility data analytics, applications of deep-learning/machine-learning methodologies in transportation domain, cooperative intelligent transportation system, and connected and automated vehicles.